\documentclass[twocolumn,prl,showpacs]{revtex4}
\usepackage{epsfig}
\begin{document}

\newcommand{\be}{\begin{equation}}
\newcommand{\ee}{\end{equation}}
\newcommand{\ber}{\begin{eqnarray}}
\newcommand{\eer}{\end{eqnarray}}

\title{
Sequence Space Localization in the Immune System Response to
Vaccination and Disease
}

\author{Michael W. Deem}
\author{Ha Youn Lee}

\affiliation{
Department of Bioengineering and
Department of Physics \& Astronomy, Rice University,
Houston, TX  77005-1892
}

\begin{abstract}
We introduce a model of protein evolution to explain limitations in the
immune system response to vaccination and disease.  The phenomenon of
original antigenic sin, wherein vaccination creates memory sequences that can
\emph{increase} susceptibility to future exposures to the same disease, is
explained as stemming from localization of the immune system response in
antibody sequence space.  This localization is a result of the roughness in
sequence space of the evolved antibody affinity constant for antigen
and is observed for diseases with high year-to-year mutation rates,
such as influenza.

\end{abstract}

\pacs{87.10.+e, 87.15.Aa, 87.17.-d, 87.23.Kg}

\maketitle

\section{Introduction}

Our immune system protects us against death by infection
\cite{Perelson1997}.
A major component of the immune system is generation of antibodies, protein
molecules that bind specific antigens.  To recognize
invading pathogens,
the immune system performs a search of the amino acid sequence space
of possible antibodies.  To find useful antibodies in the effectively infinite
protein sequence space, the immune system has evolved a hierarchical strategy.
The first step involves creating
the DNA sequences for B cells that code for
moderately effective antibodies through 
rearrangement of immune-system-specific
gene fragments from the genome \cite{Tonegawa}.
This process is called VDJ recombination.
This combinatorial process can produce on the order of $10^{14}$ different
antibodies through recombination of pieces of antibodies.
The second step, which occurs when a specific antigen invades our body,
is somatic hypermutation.
Somatic hypermutation is the process of mutation that occurs when the
B cells that produce the antibodies divide and multiply.
Only those B cells that produce antibodies that bind the antigen with
higher affinity are propagated by this mutation and selection
process, and another name for this process is affinity
maturation.  Somatic hypermutation is essentially a search of the
amino acid sequence space at the level of individual point mutations
\cite{Griffiths}.

The consequence of an immune system response to antigen is the 
establishment of a state of memory \cite{Gray}.
Immunological memory is the ability of the immune system to respond more
rapidly
and effectively to antigens that have been encountered previously. Specific
memory is maintained in the DNA of long-lived memory B cells that can persist 
without residual antigen.

 Although our immune system is highly effective, some limitations have been
reported. The phenomenon known as ``original antigenic sin'' is the
tendency for antibodies produced in response to exposure to 
influenza virus antigens to
suppress the creation of new, different antibodies in response to exposure to
different versions of the flu \cite{Fazekas1}.
Roughly speaking, the immune system responds only to the antigen
fragments, or epitopes, that are in common with the original flu virus.
As a result, individuals vaccinated against the flu
may become \emph{more} susceptible to infection 
by mutated strains of the flu than would individuals
receiving no vaccination.  The details of how original antigenic
sin works, even at a qualitative level, are unknown.

In this Letter, we offer an explanation for the reported limitations in
the immune system response using a model of protein evolution.
We describe the dynamics of affinity maturation 
by a search for increased binding constants
 between antibody and antigen
in antibody sequence space.  It is shown that 
an immune system response to an antigen generates localized memory B cell
sequences.
This set of localized sequences
reduces the ability of the
immune system to respond to subsequent exposures to different
but related antigens.
It is this competitive process between memory sequences and the VDJ
recombinations of secondary
exposure that is responsible for the reported limitations in the
immune system.

We use a random energy model to represent the interaction between
the antibodies and the influenza proteins.  This model captures
the essence of the correlated ruggedness of the interaction energy in the
variable space, the variables being the antibody amino acid sequences
and the identity of the disease proteins, and the correlations
being mainly due to the physical structure of the antibodies.  The random
energy model allows study of the sequence-level dynamics of the immune/antigen
system, which would otherwise be
an intractable problem at the atomic scale, with $10^4$ atoms
per antibody, $10^8$ antibodies per individual, $6 \times
10^9$ individuals, and many possible influenza strains.
Use of random energy theory to treat correlations in otherwise intractable
physical systems goes back at least to Bohr's random matrix theory for
nuclear cross sections \cite{Bohr1936} and has been used for
quantum chaos, disordered mesoscopic
systems, QCD, and quantum gravity \cite{Guhr1998}.
Close to the present application is the study of
spin glasses by random energy models \cite{SK1975,Derrida1980},
protein folding by coarse-grained models \cite{Wolynes1987,Gutin1993},
and evolutionary systems by NK-type models
\cite{Bogarad,Derrida1991,Weisbuch1990,Drossel2001}.

In detail, the generalized NK model we use
considers three different kinds of interactions
within an antibody: interaction within a subdomain $(U^{\rm sd})$, 
interactions between
subdomains $(U^{\rm sd-sd})$, and direct binding interaction between antibody
and antigen $(U^{\rm c})$.  In the context of protein evolution,
parameters of the model have been calibrated \cite{Bogarad,Kauffman,Perelson}.  
The energy function of a protein is given by
\begin{equation}
U=\sum_{i=1}^{M} U_{\alpha_{i}}^{\rm sd} 
+\sum_{i>j=1}^{M} U_{ij}^{\rm sd-sd} 
+\sum_{i=1}^{P} U_{i}^{\rm c} \ , 
\end{equation}
where $M$ is the number of antibody secondary structural subdomains,
and $P$ is the number of antibody
amino acids contributing directly to the binding. 
The subdomain energy $U^{\rm sd}$ is 
\begin{equation}
U_{\alpha_{i}}^{\rm sd}= {1 \over \sqrt{M (N-K+1) } } \sum_{j=1}^{N-K+1}
\sigma_{\alpha_i} (a_j, a_{j+1}, \cdot \cdot \cdot, a_{j+K-1}),
\end{equation}
where $N$ is the number of amino acids in a subdomain, 
and $K$ is the range of local interaction within a subdomain.
All subdomains belong to one of $L=5$ different types
(\emph{e.g.}, helices, strands, loops, turns, and others).
The quenched Gaussian random number $\sigma_{\alpha_i}$ is different
for each value of its argument for a given subdomain type, $\alpha_i$.
All of the Gaussian $\sigma$ values have zero mean and unit variance.
The energy of interaction between secondary structures is
\begin{eqnarray}
U_{i j}^{\rm sd-sd}= \sqrt{2 \over D M (M-1)}
\sum_{k=1}^{D} && \sigma^{k}_{i j} (a^{i}_{j_1}, 
\cdot \cdot \cdot, a^{i}_{j_{K/2}}; \nonumber \\
&& \hspace*{0.8cm} a^{j}_{j_{K/2+1}},
\cdot \cdot \cdot, a^{j}_{j_K}) \ .
\nonumber \\ 
\end{eqnarray}
We set the number of interactions between secondary structures 
at $D=6$.
 Here $\sigma^{k}_{i j}$ and the interacting
amino acids, ${j_1, \cdot \cdot \cdot, j_K}$, are selected 
at random for each interaction $(k, i, j)$.
The chemical binding energy of each antibody amino acid to the antigen
is given by 
$U_i^{\rm c}= \sigma_i (a_i) / \sqrt {P} $.
The contributing amino acid, $i$, and the 
unit-normal weight of the binding, $\sigma_i$, are chosen at random.
Using experimental results, we take
$P=5$ amino acids to contribute directly 
to the binding event.
Here we consider only five chemically distinct amino acid classes
(\emph{e.g.}, negative, positive, polar, hydrophobic, and other)
since each different type of amino acid behaves as a completely different
chemical entity within the random energy model.

The generalized NK model, while a simplified description of real proteins,
captures much of the thermodynamics of protein folding and ligand binding.
In the model, a specific B cell repertoire is represented by a specific
set of amino acid sequences.
Moreover, a specific instance of the random parameters
within the model represents a specific antigen.
An immune response that finds
a B cell that produces an antibody with high affinity constant
to a specific antigen corresponds in the model
to finding a sequence having a low energy for a specific 
parameter set.

The random character of the generalized NK model makes the
energy rugged in antibody sequence space.  The energy is,
moreover, correlated by the local antibody structure ($K = 4$),
the secondary antibody structure ($U^{\rm sd-sd}$),
and the interaction with the influenza proteins ($U^{\rm c}$).
As the immune system explores the space of possible antibodies,
localization is possible if the correlated ruggedness of the
interaction energy is sufficiently great.

Since the variable region in each light and heavy chain of an
antibody is about 100 amino acids long, and most of the binding
occurs in the heavy chain, we choose a sequence length of 100.
We choose $M=10$ since there are roughly 10
secondary structures in a typical antibody and thus choose $N=10$.
The immune system contains of the order of $10^8$ B cells divided into
different specificities, and
the frequency of a specific B cell participating in
the initial immune response is roughly
1 in $10^5$ \cite{Janeway}.
Hence, we use $10^3$ sequences during an immune response.

The hierarchical strategy of the immune system is used
to search the antibody sequence space for high affinity antibodies.
Initial combination of optimized subdomains is followed by a
point mutation and selection procedure \cite{Bogarad}.
To mimic combinatorial joining of gene segments during B cell development,
we produce a naive B cell repertoire by
choosing each subdomain sequence from pools
that have $N_{\rm pool}$ amino acid segments
obtained by minimizing the appropriate $U^{\rm sd}$.
To fit the theoretical heavy-chain diversity of
$3 \times 10^{11}$ \cite{Janeway}, 
we choose $N_{\rm pool} =  3$ sequences among the top 300
sequences for each subdomain type.

Somatic hypermutation
occurs at the rate of roughly one mutation per variable regions of light and
heavy chains per cell division, which occurs
every 6 to 8 hours during intense cell proliferation \cite{French}.
Hence, in our simulation, we do 0.5 point mutations per sequence, keep
the best (highest affinity) $x=20$\% sequences, and then 
amplify these back up to a total of
$10^3$ copies in one round, which corresponds to $1/3$ day.
That is, the probability of picking one of the, possibly mutated, sequences 
for the next round is
$p_{\rm select} = 1/200$ for $ U \le U_{200}$ and
$p_{\rm select} = 0 $ for $    U > U_{200}$,
where $U_{200}$ is the 200th best energy of the $10^3$ sequences after
the mutation events, and
this equation is employed $10^3$ times to select randomly the $10^3$ sequences
for the next round.
Given a specific antigen, \emph{i.e.}\ a specific
set of interaction parameters, we do
 30 rounds (10 days) of point
mutation and selection in one immune response. In this way, memory B cells
for the antigen are generated. 
 
 The affinity constant is given as a function of energy,
\begin{equation}
K^{\rm eq}= \exp(a-b U) \ ,
\label{eq:affinity}
\end{equation}
where $a$ and $b$ are determined by the dynamics of the
mutation and selection process.
Affinity constants resulting from
VDJ recombination are roughly
$10^4$, affinity constants after the first response of
affinity maturation are roughly
$10^6$, and affinity constants after a second response of
affinity maturation are roughly
$10^7$ \cite{Janeway} (values that fix the selection strength, $x$).
By comparison to the dynamics of the model, 
we obtain $a=-18.56, b=1.67$.

The memory B cells, key to immunological memory, give
rapid and effective response to the same antigen due to their
increased affinity for previously-encountered antigens.
We focus on the role of the memory cells in the immune response 
to different antigens.
The distance between a first antigen and the second antigen 
is given by the probability, $p$, of 
changing parameters of interaction within subdomain ($U^{\rm sd}$), 
subdomain-subdomain ($U^{\rm sd-sd}$),
and chemical binding ($U^{\rm c}$) terms.
Within $U^{\rm sd}$, we change only the subdomain type, $\alpha_i$,
not the
parameters $\sigma_\alpha$, which are probably
fixed by structural biology
and should be independent of the antigen.

We estimate the number of memory and naive B cells 
that participate in the immune response to the second antigen by
the ratio of the respective affinity constants.
From the definition of the affinity constant,
$K^{\rm eq}= \left [ {\rm Antigen:Antibody} \right ]  / \left\{
{\left[ {\rm Antigen} \right] 
\left [ {\rm Antibody} \right ]} \right\}$,
the binding probability is proportional to the affinity constant
and to the concentration of antigen-specific antibody,
which is $10^2$ times greater for the memory sequences \cite{Janeway}.
We measure the average affinity for the second antigen
of the $10^3$ memory cells,  $K^{\rm eq}_{\rm m}$,
and that of the $10^3$ B cells from the naive repertoire
of optimized subdomain sequences,
$K^{\rm eq}_{\rm n}$.
The ratio $10^2 K^{\rm eq}_{\rm m} / K^{\rm eq}_{\rm n}$ 
gives the ratio of memory cells to naive cells.
For exposure to the second antigen, 
we perform 30 rounds (10 days) of point mutation
and selection, starting with
$10^5 K^{\rm eq}_{\rm m} / (10^2 K^{\rm eq}_{\rm m}+K^{\rm eq}_{\rm n}) $ 
memory cells and
$10^3 K^{\rm eq}_{\rm n} / (10^2 K^{\rm eq}_{\rm m}+K^{\rm eq}_{\rm n})$
naive cells, since both memory and naive sequences participate
in the secondary response \cite{Berek}.

\begin{figure}
\begin{center}
\epsfig{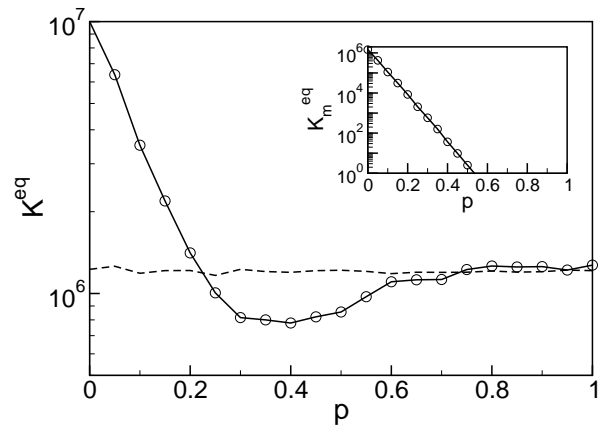}
\end{center}
\vspace{0.5cm}
\caption[]{
The evolved affinity constant to a second antigen after exposure to
an original antigen that differs by probability $p$ (solid line).
The dotted line represents the affinity constant without previous exposure.
The affinity constant is generated by exponentiating,
as in Eq.\ (\ref{eq:affinity}),
the average of the best binding energy, using
5000 instances of the model.  In inset is shown the affinity of
the memory sequences for the mutated antigen.
}
\label{fig:localization}
\end{figure}

Figure \ref{fig:localization} shows the evolved affinity constant
to a second antigen if the exposure to a first antigen 
exists (solid line) or not (dashed line) as a function of the difference
between the first and second antigen, $p$, or
``antigenic distance'' \cite{Smith}.
When the difference is small, the exposure to a first antigen
leads to higher affinity constant than without exposure, which is
why immune system memory and
vaccination is effective.
For a large difference, the antigen encountered in the first exposure
is uncorrelated with that in the second exposure, and so immune system 
memory does not play a role.
Interestingly, the immunological memory from the first exposure
actually gives worse protection, \emph{i.e.}\  a
lower affinity constant, for intermediate differences---which is
original antigenic sin.

The dynamics of the immune response (Fig.\ \ref{fig:localization})
depend upon the constants of Nature, \emph{i.e.}\
the parameters of the model.
For example, in an organism with a smaller immune system, such as the mouse,
there are fewer starting sequences and
less favorable binding constants are measured in the same
number of rounds: a factor of 0.5 reduction
in the number of starting sequences leads to a 0.64 reduction in the evolved
binding constant, but a similar degree of original antigenic sin as in
Fig.\ \ref{fig:localization}.  On the other hand, if more
rounds are performed, better binding constants are found in the primary and
secondary responses, but the secondary response is not as improved over
the primary response as when
using 30 rounds, because the evolved sequences are becoming
more localized in ever deeper wells.  The degree of original antigenic sin is,
however, similar in the range of 30 to 60 rounds per response.  
Similarly, if the roughness of the
energy upon disease mutation is increased,
for example by assuming that mutation of the influenza actually changes
the $\sigma_\alpha$, the degree of original antigenic sin increases
substantially, by a factor of 2, because the barriers between the
regions of localization in sequence space are increased.
On the other hand, if the 
concentration of the memory cells is decreased, the contribution of the
memory cells to the dynamics is reduced, and the original antigenic
sin phenomenon decreases, almost disappearing when the memory and
naive antibody concentrations are equal.
The average number of mutations leading to the best antibody,
a measure of the localization length when original antigenic sin
occurs, is 15 for the first response and rises from 5 to 15
for the second response in the range $0 \le p < 0.30$.
Interestingly, the average number of mutations rises slightly above 15
in the range $0.30 \le p < 0.70$, indicating that in the original
antigenic sin region
more mutations are necessary for the compressed ensemble
of memory sequences from the primary response to evolve to a suitable
new state in the secondary response.
Larger selection strengths ($x < 20$\%) cause
more localization and original antigenic sin, in shallower wells for
small $x$, and smaller strengths ($x > 20$\%) lead to less 
evolution.

For small values of $p$, $p < 0.19$, the memory B cells produce antibodies
with higher affinities, $K^{\rm eq}_{\rm m} > 10^4$,
 for the new antigen than do naive B cells.
 The binding constant of the memory antibodies steadily
decreases with $p$, reaching the non-specific value of 
$K^{\rm eq}_{\rm m} = 10^2$ at $p = 0.36$
(see inset in Fig.\  \ref{fig:localization}), which, interestingly,
is less than the range to which 
original antigenic sin extends,
$0.23 < p < 0.60$.
 These model predictions are in good
agreement with experimental data
on cross-reactivity, which ceases to occur
when the amino acid sequences are more than 33--42\% different
\cite{East}.

\begin{figure}
\begin{center}
\epsfig{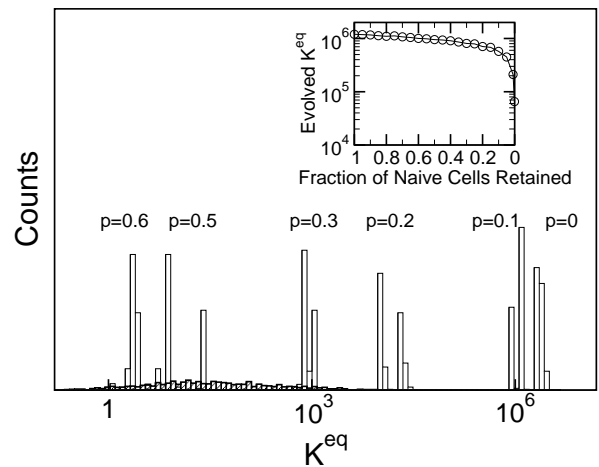}
\end{center}
\vspace{0.5cm}
\caption[]{
The affinity distribution of antibodies from
naive B cells (hatched) or memory cells
(open) for the antigen of second exposure.
In the inset is shown the binding constant that evolves in
the primary response when an initial selection step retains
only the top fraction of the naive cells.
}
\label{fig:histogram}
\end{figure}

The ineffectiveness of immune system memory over a window of $p$ values
can be understood from the localization of memory B cell sequences.
Figure \ref{fig:histogram} displays distributions 
of memory and naive affinity constants
for the second antigen.
Notice that the memory sequences are highly homogeneous and lack diversity
compared to the naive sequences.  Indeed, original antigenic sin
arises mainly because the memory sequences from the primary response
suppress use of naive sequences in the tail of the distribution for the
secondary response.  Although those naive sequences initially look
unpromising, they may actually evolve to sequences
with superior binding constants.  
The inset to Fig.\ \ref{fig:histogram} 
illustrates this phenomenon.
Interestingly, when half of the
distribution is removed, the reduction in the binding constant is just about
that which occurs in original antigenic sin, Fig.\  \ref{fig:localization}.
%
%


In summary, the generalized NK model is shown to successfully
model immune system
dynamics.  A localization mechanism for the original antigenic sin phenomenon 
observed in the flu is explained.
Localization of antibodies in the amino acid sequence space around
memory B cell sequences is shown to lead to a decreased ability of the
immune system to respond to diseases with year-to-year mutation rates
within a critical window.
This localization occurs because of the ruggedness of the evolved
affinity constant in amino acid sequence space.
From the model dynamics, we find that
memory sequences can both outcompete the non-vaccinated
immune response and become trapped in local minima.
Memory sequences with affinity constants 
initially superior to those from 
naive sequences can be selected by the dynamics,
and these memory sequences can lead to poorer evolved affinity constants, 
to the detriment of the immune system for intermediate disease
mutation rates. 
These results suggest several implications for vaccination strategy:
the difference between vaccinations administered on a repeated basis should
be as great as practicable, and suppression of the memory B cell response 
may be helpful during vaccination against highly variable antigens.

M.W.D. thanks Jeong-Man Park for stimulating discussion.
This research was supported by the National Science and 
Camille \& Henry Dreyfus Foundations.

Corresponding e-mail: mwdeem@rice.edu.
\hbox{}\vspace{-0.25in}

\bibliography{immune}

\end{document}